\documentclass[fleqn,usenatbib,useAMS]{mnras}

\usepackage{newtxtext,newtxmath}

\usepackage[T1]{fontenc}
\usepackage{ae,aecompl}

\DeclareRobustCommand{\VAN}[3]{#2}
\let\VANthebibliography\thebibliography
\def\thebibliography{\DeclareRobustCommand{\VAN}[3]{##3}\VANthebibliography}

\usepackage{graphicx}	
\usepackage{amsmath}	
\usepackage{amssymb}	
\usepackage[para,online,flushleft]{threeparttable}

\newcommand{\hii}{H~II~}
\newcommand{\hiirs}{\hii~regions~}
\newcommand{\uchii}{UC~\hii~}
\newcommand{\uchiir}{UC~\hii~region~}
\newcommand{\uchiirs}{UC~\hii~regions~}
\newcommand{\uchiie}{UC~\hii+EE~}
\newcommand{\uchiies}{UC~\hii+EE~regions~}
\newcommand{\kms}{~km~s$^{-1}$}

\title[The ultracompact \hii region with extended emission G43.89--0.78]{Ultracompact HII regions with extended emission: The case of G43.89--0.78 and its molecular environment}

\author[Eduardo de la Fuente et al.]{
Eduardo de la Fuente,$^{1}$\thanks{Corresponding author e--mail: eduardo.delafuente@academicos.udg.mx }\thanks{Also HAWC Gamma--Ray National Laboratory (CONACyT), M\'exico. Research visit at Institute for Cosmic Ray Research (Feb. 2020), University of Tokyo, Kashiwa Campus, Japan. }
Daniel Tafoya$^{2}$, 
Miguel A. Trinidad$^{3}$,
Alicia Porras$^{4}$,
\newauthor
Alberto Nigoche-Netro$^{5}$,
Simon N. Kemp$^{5}$,
Stanley E. Kurtz$^{6}$,
Jos\'e Franco,$^{7}$ 
\newauthor
and Carlos A. Rodr\'iguez--Rico$^{3}$
\\
$^{1}$Departamento de F\'{i}sica, CUCEI, Universidad de Guadalajara, Blvd. Gral. Marcelino Garc\'{i}a Barrag\'an 1421, Ol\'impica, 44430, Guadalajara, Jalisco, M\'exico.
\\
$^{2}$Department of Space, Earth and Environment, Chalmers University of Technology, Onsala Space Observatory, 439~92 Onsala, Sweden.
\\
$^{3}$Departamento de Astronom\'{i}a, Universidad de Guanajuato, Apartado Postal 144, 36000, Guanajuato, Guanajuato, M\'exico.
\\
$^{4}$Instituto Nacional de Astrof\'{i}sica, \'Optica y Electr\'onica, Luis E. Erro N\'umero 1, 72840,  Tonantzintla, San Andr\'es Cholula, Puebla, M\'exico.
\\
$^{5}$Instituto de Astronom\'{i}a y Meteorolog\'{i}a, CUCEI, Universidad de Guadalajara, Av. Vallarta 2602, 44130, Guadalajara, Jalisco, M\'exico.
\\
$^{6}$Instituto de Radioastronom\'{i}a y Astrof\'{i}sica, UNAM, Antigua Carretera a P\'atzcuaro Numero 8701,  58089, Morelia, Michoac\'an, M\'exico.
\\
$^{7}$Instituto de Astronom\'{i}a, UNAM, Apartado Postal 70--264, CDMX, 04510, M\'exico
}

{\bf \date{To be Published on Monthly Notices of the Royal Astronomical Society Main Journal.\\ Accepted 2020 July 15. Received 2020 July 15; in original form 2020 June 10}}

\pubyear{2015}

\begin{document}
\label{firstpage}
\pagerange{\pageref{firstpage}--\pageref{lastpage}}
\maketitle


\begin{abstract}

The Karl Jansky Very Large Array (VLA), Owens Valley Radio Observatory (OVRO), Atacama Large Millimetric Array (ALMA), and the infrared \textit{Spitzer} observatories, are powerful facilities to study massive star formation regions and related objects such as ultra--compact (UC) \hii regions, molecular clumps, and cores. We used these telescopes to study the \uchiir G43.89--0.78. The morphological study at arcminute scales using NVSS and \textit{Spitzer} data shows that this region is similar to those observed in the \textit{ bubble--like} structures revealed by \textit{Spitzer} observations. With this result, and including a physical characterization based on 3.6 cm data, we suggest G43.89--0.78 be classified as an \uchiir with Extended Emission because it meets the operational definition given in this paper comparing radio continuum data at 3.6 and 20~cm. For the ultra-compact component, we use VLA data to obtain physical parameters at 3.6~cm confirming this region as an \uchii region. Using ALMA observations, we detect the presence of a dense ($2.6\times10^7$ cm$^{-3}$) and small ($\sim$ 2.0\arcsec; 0.08 pc) molecular clump with a mass of 220 M$_{\odot}$ and average kinetic temperature of 21~K, located near to the \uchii region. In this clump, catalogued as G43.890--0.784, water masers also exist, possibly tracing a bipolar outflow. We discover in this vicinity two additional clumps which we label as G43.899--0.786 (T$_d$ = 50 K; M = 11 M$_{\odot}$), and G43.888--0.787 (T$_d$ = 50 K; M = 15 M$_{\odot}$).

\end{abstract}




\section{Introduction}
\label{sec:sec1}

The study of \uchiirs is very important to understand the high--mass star formation process because they trace regions with recently formed ionizing (OB-type) stars. The \uchiirs are often related to hot molecular cores and maser emission \citep[e.g,][and references therein]{Hofner1996,Garay1999,Kurtz2000b,Churchwell2002,delaFuente2018}. They were identified by \citet{Wood1989} and \citet{Kurtz1994} as small (sizes $\leq$ 0.1 pc) and dense ($\gtrsim$ 10$^4$ cm$^{-3}$) \hii regions with high emission measures ($\geq$~10$^7$ {${\rm pc\ cm}^{-6}$}). The physical parameters of the ionized gas are determined by radio continuum (RC) and radio recombination line (RRL) observations \citep[e.g,][]{Oster1961,Mezger1967,Schraml1969,Dupree1970,Panagia1978}.

There are five {\it booms} in the history of the topic: 1.- the original studies via single-dish observations \citep[e.g,][and references therein]{Mezger1967,Ryle1967,Israel1973}, 2.- the physical characterization for the first time via VLA interferometric observations \citep[][]{Wood1989,Kurtz1994}, 3.- observations (and studies) of {\it extended emission} (EE) at arcmin scales by \citealt{delaFuente2007} \citep[see][]{Kurtz1999,Kim2001,Kim2002,Kim2003,Ellingsen05,delaFuente2009a,delaFuente2009b,delaFuente2018,delaFuente2020}, 4.- the discovery of time variation in the radio flux density \citep[][]{Franco2004,Galvan2008} and, 5.- the detection of hypercompact (HC) \hii regions \citep[][]{Gaume1995,Sewilo2004, Sewilo2008,DePree2011}. Reviews can be found in \citet[][]{Kurtz2000a,Churchwell2002,delaFuente2009a} and references therein.

\defcitealias{delaFuente2020}{Paper~I}
In the morphological study performed by \citet[][\citetalias{delaFuente2020} hereafter]{delaFuente2020}, we found that extended radio continuum emission (EE) resembles bubble--like structures found in the {\it Bubbling Universe} revealed by {\it Spitzer} observations \citep[e.g.,][and references therein]{Churchwell2006,Churchwell2007,Churchwell2009,Everett2010,Simpson2012}. \citetalias{delaFuente2020} also shows that for all the sources of the sample, except for G12.21--0.10 (G12.21 hereafter), the {\it Spitzer} MIPS emission at 24~$\mu$m is saturated at the position of the UC component (the \uchii region). Hence, this MIPS band could be a good tracer of \uchiirs and especially those \uchii with EE (\uchiie regions) via visual inspection.  Such infrared data could thus be an important complement to morphological studies by avoiding the spatial filtering of extended emission inherent to radio--interferometers.

Water maser emission is known to be a common --- if not ubiquitous --- phenomenon in star formation regions \citep[e.g.,][and references therein]{Elitzur1992,Garay1999,Trinidad2003,Codella2004}. Nevertheless, various questions remain regarding the relation of the masers to the star formation process. \citet{Hofner1996} show that for \uchiirs with cometary morphologies, water masers are always located in clumps, near, but offset from, the cometary arc. These masers could be related to nearby, embedded young stellar objects rather than to the \hii regions themselves. 

G12.21 and G43.89--0.78 (G43.89 hereafter) are good candidates to study the nature of these water maser clumps because these two sources have larger separations between the masers and the \hii region.  Hence, the maser clump can be spatially resolved from ionized gas.  In this context, the molecular clump with water masers observed in the UC component of the \uchiie region G12.21 was characterized as a hot molecular core via NH$_{\rm 3}$ (2,2) and (4,4) VLA observations by \citet{delaFuente2018}. These authors start their analysis with the detection of $^{13}$CS(2--1) at the position of the water masers. They confirm that the masers arise within the molecular core instead of the \hii region. A similar analysis was not performed for G43.89, hence,  the origin of the masers remained unknown for this source.  The purpose of the present paper is to provide such an analysis for G43.89
and further, elucidate the origin of water masers during the star formation process.

G43.89 (IRAS 19120+0917) is a cometary \uchiir with a diameter of $\sim$4$\arcsec$ \citep{Wood1989}. Its distance is uncertain. \citet{Wink1982} via H76$\alpha$ observations ($V_{LSR} = 50.9$ \kms) find near/far kinematic distances of 3.9/10.5~kpc. Using the CS velocity of $V_{\rm LSR}$ of 54.2\kms reported by \citet{Bronfman1996} and the galactic rotation model of \citet{Wouterloot1990}, we estimate near and far kinematic distances of 6.5 and 8.0~kpc, respectively. \citet{Kuchar1990} report HI absorption up to the terminal velocity, and adopt a distance of 8.8~kpc. \citet{Araya2002} found a distance of 8.12~kpc. \citet{Wenger2018} report a Monte Carlo parallax distance of 7.82 kpc. We adopt 8.0 kpc as the distance to G43.89 for our analysis. 

The water maser emission toward G43.89 was first detected by \citet{Churchwell1990} and mapped with high spatial resolution by \citet{Hofner1996}. CH$_3$OH maser emission at 6.67 GHz was reported by \citet{Schutte1993}. \citet{Baudry1997} and \citet{Codella2010} suggest the presence of OH maser emission. The NH$_3$ (1,1), (2,2) and (3,3) observations of \citet{Codella2010} show weak emission with no distinctive morphology. CS studies were performed by \citet{Hatchell1998} and \citet{Olmi1999}. \citet{Bronfman1996} detected emission of CS($J$=2$\rightarrow$1). Single--dish C$_2$H$_3$N and NH$_3$ observations by \citet{Olmi1993} indicate a gas temperature T$\sim$30~K. The high-excitation NH$_3$ studies of \citet{Cesaroni1992} and \citet{Olmi1993} indicate the presence of hot, dense gas, but \citet{Hatchell1998, Hatchell2000} find no evidence for a hot molecular core. Single--dish CO observations by \citet{Shepherd1996} detect a total linewidth of 39.1~km s$^{-1}$ which suggests an outflow. Infrared fluxes were measured and spectra taken  by \citet{Doherty1994} and \citet{Hanson2002}.

Here, we perform a study of G43.89 similar to that of G12.21 by \citet{delaFuente2018}.  Our goals are to investigate where the water masers arise, and to characterize both the ionized and molecular gas components of G43.89. In Section~\ref{sec:sec2} we describe the observations and data reduction procedures. Results and discussion are given in Section~\ref{sec:sec3}. Summary and conclusions are given in Section~\ref{sec:summary}. The physical characterization at 3.6~cm for \uchiies of \citetalias{delaFuente2020} and the comparison with the \citet{Kim2001} study at 20~cm is shown in Appendix~ \ref{sec:appendixA}. The latter helps to confirm G43.89 as an \uchiie region.

\begin{table*}
\caption{High Resolution Radio continuum observations\tnote{a}}
\label{tab:tab1}
\begin{threeparttable}

\begin{tabular}{lcccccccr}
\hline

\, $\lambda$ & Instrument & S$_{\nu}$\tnote{b} & Beam Size & PA & Source Size\tnote{c} & rms & Observational    \\
\, (cm) &  & (mJy) & ( $\arcsec \times \arcsec$ ) & ( $^{\circ}$ )  & ($\arcsec$) & (mJy beam$^{-1}$) & Program    \\

\hline

\, 0.3 & ALMA & 401.0 &  1.90$\times$1.58 & +57.0 & 2.29 & 0.30 & 2015.1.00280.S  \\ 
\, 0.3 & OVRO & 260.0 & 3.45$\times$2.45 & --58.0 & 1.70 & 3.00 & --  \\
\, 0.7 & VLA--CnB & 396.0 & 1.46$\times$0.81 & +90.0 & 2.00  & 0.80 & AK423    \\
\, 2.0 & VLA--CnB & 420.0 & 1.58$\times$0.95 & --08.0 & 2.00 & 0.30 & AK423    \\
\, 3.6 & VLA--CnB & 570.0 & 2.86$\times$1.85 & --08.0 & 3.20 & 0.20 & AK423   \\

\hline

\end{tabular}
\begin{tablenotes}

\item[a] Technical details of the low-resolution VLA observations (the 20~cm NVSS data) are given in \citet{Condon1998} \\
\item[b] Uncertainty on integrated flux is 20\% at 0.3 and 0.7 cm; 10\% at 2~cm; 5\% at 3.6~cm; and 10\% at 20.0~cm.  \\
\item[c] Deconvolved size obtained using the task \textit{imfit} of AIPS; $\Theta_{\rm s}$ = $\sqrt{\Theta_{\rm x} \Theta_{\rm y}}$. \\

\end{tablenotes}
\end{threeparttable}
\end{table*}

\begin{table*}
\centering
\caption{Parameters of the spectral line observations}
\label{tab:tab2}
\begin{threeparttable}

\begin{tabular}{lccccccccr}
\hline

\,  Line & Instrument & Rest frequency & Channel width & rms &  Beam Size & PA \\
\,  &  & (GHz) & (km~s$^{-1}$) & (mJy beam$^{-1}$) &  ( $\arcsec \times \arcsec$ ) & ( $^{\circ}$ )  \\

\hline
\,  H40$\alpha$ & ALMA & 99.02296 & 3.0 & 1.5 & 1.68$\times$1.34 & +66  \\
\,  H41$\alpha$ & OVRO & 92.03444 & 3.3 & 25.0 & 3.45$\times$2.45 & --58    \\
\,  H42$\alpha$ & ALMA & 85.68839 & 3.0 & 2.0 & 1.92$\times$1.68 & +64 \\
\,  $^{13}$CS($J$=2$\rightarrow$1) & OVRO & 92.49431 & 1.6 & 35.0 & 3.45$\times$2.45 & --58 \\
\,  CS($J$=2$\rightarrow$1) & ALMA & 97.98095 & 3.0 & 2.0 & 1.67$\times$1.43 & +68  \\
\,  HCN($J$=1$\rightarrow$0) & ALMA & 88.63160 & 3.0 & 2.0& 1.86$\times$1.63 & +59  \\
\,  H$^{13}$CN($J$=1$\rightarrow$0) & ALMA & 86.33992 & 3.0 & 2.0 & 1.91$\times$1.65 & +62   \\
\,  HCO$^{+}$($J$=1$\rightarrow$0) & ALMA & 89.18852 & 3.0 & 2.0 & 1.85$\times$1.63 & +60  \\
\,  SiO($J$=2$\rightarrow$1) & ALMA & 86.84696 & 3.0 & 2.0  & 1.89$\times$1.65 & +62 \\
\hline

\end{tabular}
\begin{tablenotes}
\end{tablenotes}
\end{threeparttable}
\end{table*}

\section{Observations}
\label{sec:sec2}

\subsection{{\it Spitzer} and low--resolution VLA Observations}
\label{sec:sec2.1}

Infrared observations were taken from the {\it Spitzer} telescope \citep{Werner2004} Legacy Programs.  IRAC \citep[][]{Fazio2004} data from the GLIMPSE Spitzer and Ancillary Data \citep{Benjamin2003} and MIPS \citep[][]{Rieke2004} 24$\mu$m data from MIPSGAL \citep[][]{Carey2009} were retrieved. 

The IR imagery traces extended emission of \uchiirs at arcmin scales. The stellar content and PAHs at 3.3~$\mu$m are observed in the [3.6]~$\mu$m IRAC1 band. The ionized gas is detected in the [4.5]~$\mu$m IRAC2 band. The [5.8]~$\mu$m IRAC3 band traces cold dust and PAHs at 6.2~$\mu$m. The [8.0]~$\mu$m IRAC4 band traces PAHs where the 7.7~$\mu$m emission is predominant. The MIPS [24]~$\mu$m band detects warm dust and embedded YSOs as point--like sources. More details and a description of characteristics in these bands in \uchiies is presented in \citetalias{delaFuente2020} (and references therein) \citep[see also][]{Robitaille2006,Robitaille2008,delaFuente2009a,delaFuente2009b}.

Because there are no VLA observations at 3.6~cm in configurations C or D for G43.89, we follow \citet{Kurtz1999} and use 20~cm NVSS observations from \citet[][]{Condon1998} to trace the EE. NVSS has a sensitivity to structures 7\arcmin--15\arcmin, and the EE seems to be associated with IR {\it Spitzer} bubbles (see \citetalias{delaFuente2020}), with the NVSS emission encompassing the arcminute-scale IR structures.

\subsection{High resolution observations}
\label{sec:sec2.2}

\subsubsection{VLA and OVRO observations}
\label{sec:sec2.2.1}

VLA observations of RC at 0.7, 2, and 3.6~cm toward G43.89 were performed simultaneously with those of G12.21, reported by \citet{delaFuente2018}. In a similar way to G12.21, we used the Owens Valley Radio Observatory (OVRO) Millimeter Array to observe RC emission at 0.3 cm, and the $^{13}$CS($J$=2$\rightarrow$1) ($\nu_0 = 92.49430$ GHz) line and the H41$\alpha$ ($\nu_0 = 92.03445$ GHz) line emission toward G43.89. The observations were carried out in March and April, 1996, using both equatorial and high--resolution configurations. Details of these observations, instrumental set--up, calibration and data reduction procedures are presented in \citet{delaFuente2018}. 

Calibration and standard data reduction procedures were performed using the AIPS package\footnote{AIPS is produced and maintained by the National Radio Astronomy Observatory, a facility of the National 
Science Foundation operated under cooperative agreement by Associated Universities, Inc.}. Observational parameters for the high-resolution RC are shown in Table~\ref{tab:tab1}. Table~\ref{tab:tab2} gives the corresponding parameters for the spectral line observations. 

\subsubsection{ALMA observations}
\label{sec:sec2.2.2}

We retrieved archival Atacama Large Millimetre Array (ALMA) data of the project 2015.1.00280.S (PI: R. Cesaroni) to study the molecular, mm--continuum, and RRL emission associated with G43.89. The observations were performed using Band 3 (84-116~GHz) receivers and include two Executions Blocks (EBs) obtained on the 16th and 17th of March, 2016, with 37 and 40 12-m antennas, respectively. 

A total of six spectral windows with different bandwidths and spectral resolutions were used to target the continuum and line emission.

The highest and lowest spectral resolution of the observations are 0.42~km~s$^{-1}$ and 1.48~km~s$^{-1}$, respectively. The total observation time on G43.89 was 35 minutes (including both EBs). The baselines of the array covered a range from 15.1~m to 460.0~m, which provided an angular resolution of $\sim$1$\rlap{.}^{\prime\prime}$5 and a maximum recoverable scale of $\sim$25$^{\prime\prime}$. 

Precipitable water vapour levels ranged from 2.3 to 2.9~mm and the typical system temperature was 50--100~K for different antennas. The data were calibrated with the ALMA pipeline (36252--Pipeline-Cycle3-R4-B; CASA version 4.5.2 r36115) using J1751+0939 ($\sim$2.72~Jy at 86.751~GHz) as a flux/bandpass calibrator and J1922+1530 $\sim$260~mJy as a gain calibrator. 

Images were created using Briggs weighting with the robust parameter set to $-$0.5 in order to improve the spatial resolution, at the cost of slightly higher noise. The 0.3 cm continuum image has an average rms of 0.3~mJy~beam$^{-1}$. The corresponding molecular  line observational parameters are summarized in Table~\ref{tab:tab2}.

\begin{figure*}
\centering
    \includegraphics[width=\textwidth]{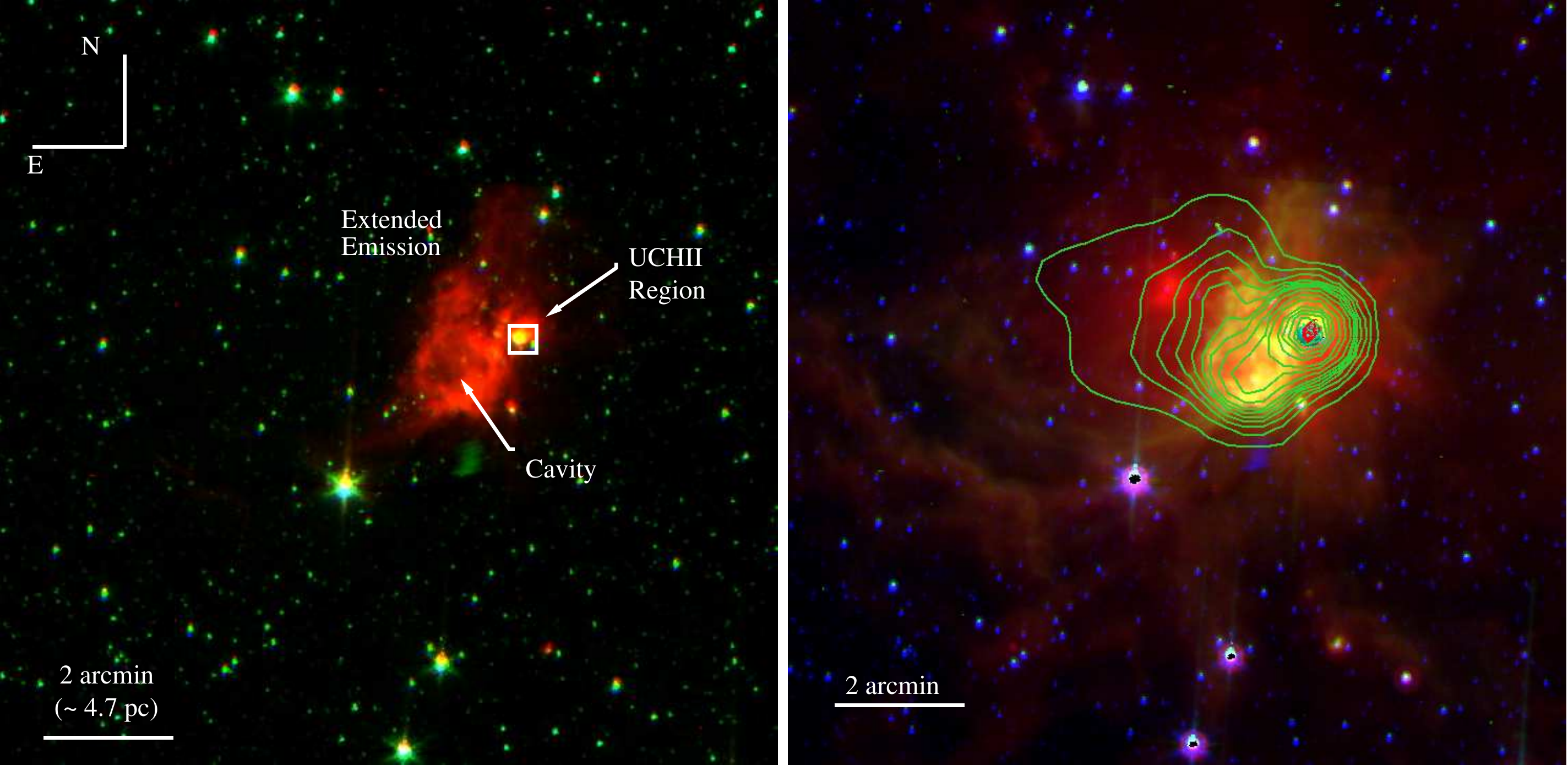}
        \caption{G43.89--0.78. {\bf Left}: {\it Spitzer} IRAC RGB (8$\mu$m + 4.5$\mu$m + 3.6$\mu$m) image. The red emission traces the extended emission. The square shows the position of the \uchii region.  A cavity--like structure is marked. The yellow emission at the position of the \uchiir is from the intense 4.5 $\mu$m emission. This emission is clearly observed in its corresponding IRAC image, and matches very well with the red emission. {\bf Right}: {\it Spitzer} IRAC+MIPS RGB (24$\mu$m + 8$\mu$m + 4.5$\mu$m) image. Green contours are from NVSS showing the 20~cm emission tracing extended ionized gas. The levels are --4, 4, 8, 16, 32... times 1.20 mJy beam$^{-1}$. The NVSS peak emission, at the position of the \uchiir, is coincident with the emission at 3.6~cm (red contours). In this small region, also 24~$\mu$m emission is saturated (cyan colour). The emission at 4.5 $\mu$m and 8~$\mu$m, shown here in yellow, matches with higher values of NVSS contours. The NVSS emission has two peaks, one related with the UC component (peak--1), and other related with the cavity (peak--2). At a distance of 8 kpc, 2\arcmin corresponds to $\sim$ 4.7 pc.}
    \label{fig:fig1}
\end{figure*}

\section{Results and discussion}
\label{sec:sec3}

We study the EE associated with G43.89 via \textit{Spitzer} and VLA low--resolution observations.  The need for this study was evident upon visual inspection of the MIPS image, which shows significant structure around the saturated \uchiir.

\subsection{The extended emission}
\label{sec:sec3.1}

IRAC and IRAC$+$MIPS RGB images for G43.89 are shown in Fig.~\ref{fig:fig1}. The IRAC morphology is similar to other \uchiies shown by \citet{delaFuente2009a,delaFuente2009b} with the EE marked by the red emission at 8.0~$\mu$m. In G43.89, the emission at 4.5~$\mu$m is considerable and matches the south--western edge of the 8~$\mu$m morphology (in red), being more intense at the position of the \uchiir (in yellow) marked with a square in Fig.~\ref{fig:fig1} (left). An almost circular structure with size $\sim$~1\arcmin, considered as a cavity, is seen at 8~$\mu$m and delineates a bubble--like shape that is characteristic of some star forming regions in the Milky Way \citep{Churchwell2006} as revealed by {\it Spitzer} observations.

\begin{table*}
\caption{Results for the ionized gas}
\label{tab:tab3}
\begin{threeparttable}
\begin{tabular}{lcccccccccc}
\hline

\, \hii & S$_{\nu}$ & Distance & Source & Source  & T$_{\rm e}$\tnote{a} & EM & n$_{\rm e}$ & M$_{\rm \hii}$  & Log(N$'_{\rm c}$) & \, Spectral\tnote{b} \\

\, Component & (mJy) & (kpc) & size (\arcsec)  & size (pc)  & (K)  & (cm$^{-6}$ pc) & (cm$^{-3}$) & M$_{\odot}$ & (s$^{-1}$) & Type (radio)    \\

\hline

\, Extended Emission & 318\tnote{c} & 8.0 & 45.0 & 1.75 & 10000 & 8.7$\times$10$^4$ & 2.2$\times$10$^2$ & 16.00 & 48.10 & O9    \\

\, Ultracompact  & 570\tnote{d} & 8.0 & 3.2 & 0.12 & 10700 & 3.7$\times$10$^7$ & 1.8$\times$10$^4$ &  0.43  & 48.44 & O8     \\

\hline
\end{tabular}
\begin{tablenotes}

\item[a] We adopt the canonical value of 1 $\times$ 10$^4$ K for the EE, and for the UC we use the electron temperature derived from the H41$\alpha$ emission using equation \ref{eqn:eqn1} with values from table~\ref{tab:tab2}). \\
\item[b] Using \citet{Panagia1973}. All sources considered at ZAMS. \\
\item[c] This flux corresponds to the NVSS peak--1 only. \\
\item[d] This flux corresponds to the VLA CnB conf. at 3.6 cm  \\

\end{tablenotes}
\end{threeparttable}
\end{table*}

\begin{table}
\centering
\caption{Results of the RRL observations}
\label{tab:tab4}
\begin{threeparttable}

\begin{tabular}{lccccc}
\hline

\,  Line  & S$_{\nu}$\tnote{a} & Size\tnote{b} &  V$_{\rm LSR}$ & $\Delta$V (FWHM) & T$_{\rm L}$ / T$_{\rm C}$   \\
\,  & (mJy) & ($\arcsec$)  & (km s$^{-1}$) & (km s$^{-1}$) & (at 0.3 cm)   \\

\hline

\,  H40$\alpha$  & 386.0 & 2.2 &  55.2$\pm$0.1 & 24.4$\pm$0.2 & 0.95$\pm$0.01    \\
\,  H41$\alpha$  & 246.0 & 3.0  & 56.6$\pm$0.4 & 25.8$\pm$0.9 & 0.95$\pm$0.02  \\
\,  H42$\alpha$  & 378.0 & 2.4 & 55.3$\pm$0.1 & 24.3$\pm$0.3 & 0.95$\pm$0.01  \\
\hline

\end{tabular}
\begin{tablenotes}
\item[a] Uncertainties on integrated flux are 20\%. Typical OVRO beam sizes for the 0.3 cm lines are 3$''$. Uncertainties on integrated flux for ALMA observations are 5\%. \\
\item[b] Deconvolved size obtained from observations using task \textit{imfit} of AIPS. $\Theta_{\rm s}$ = $\sqrt{\Theta_{\rm x} \Theta_{\rm y}}$  \\

\end{tablenotes}
\end{threeparttable}
\end{table}

\begin{figure*}
	\includegraphics[width=2\columnwidth]{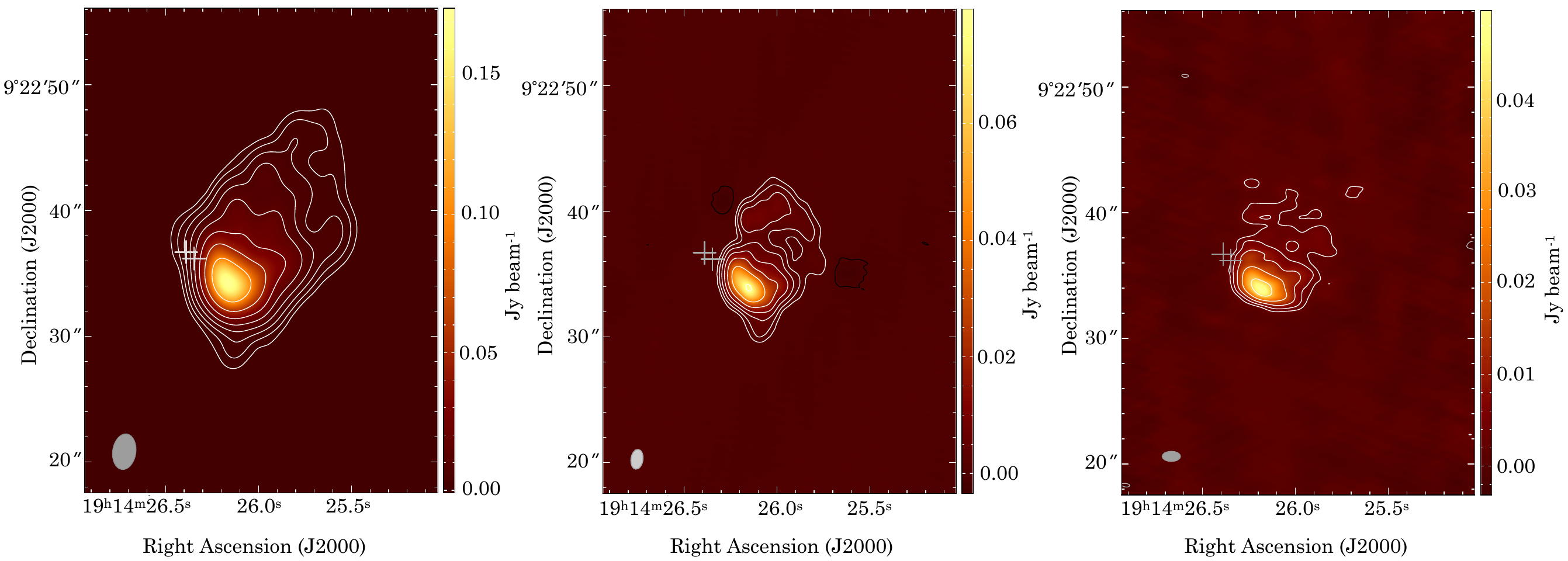}
    \caption{Radio continuum emission for G43.89--0.78 at 3.6, 2, and 0.7 cm observed with the VLA at high--resolution. The contours increase as $\pm$4$\times$rms$\times$2$^{i}$ with $i$=0,1,2,3$\ldots$ (black contours have negative values) and the rms is listed in Table~\ref{tab:tab1}. The synthesized beams (see Table~\ref{tab:tab1}) are shown in the bottom-left corner of each panel. The crosses mark the position of the water masers from \citet{Hofner1996}. These maps trace only the UC component.}
    \label{fig:fig2}
\end{figure*}

The IRAC$+$MIPS image shows the same behaviour observed in \uchiies from \citetalias{delaFuente2020}, showing a remarkable similarity with G37.55--0.11. In both sources, low--resolution RC contours coincide with the 24~$\mu$m emission; a well--traced cavity is evident at 8~$\mu$m, which is filled with 24~$\mu$m emission, and red, point--like sources, YSOs or dust condensations, are located inside. Red contours, coinciding with the saturation at 24~$\mu$m (cyan color), are from the high--resolution VLA 3.6~cm image, also shown in Fig.~\ref{fig:fig2}. This emission traces the UC component.

The NVSS emission has two peaks, one coinciding with the \uchiir position (peak--1), the other coinciding with the IRAC \textit{cavity} (peak--2). Weaker RC emission extends to the NE as part of a larger structure of ionized gas. It is noticeable that at larger scales, the RC emission matches with 24~$\mu$m emission but does not have an IRAC counterpart (such behaviour is also seen in G37.55--0.11). The IRAC emission at 4.5~$\mu$m in peak--1 matches with the yellow emission on the IRAC$+$MIPS image, supporting the presence of ionized gas at several scales from the \uchii regions.

\begin{figure*}
\centering
	\includegraphics[width=\textwidth]{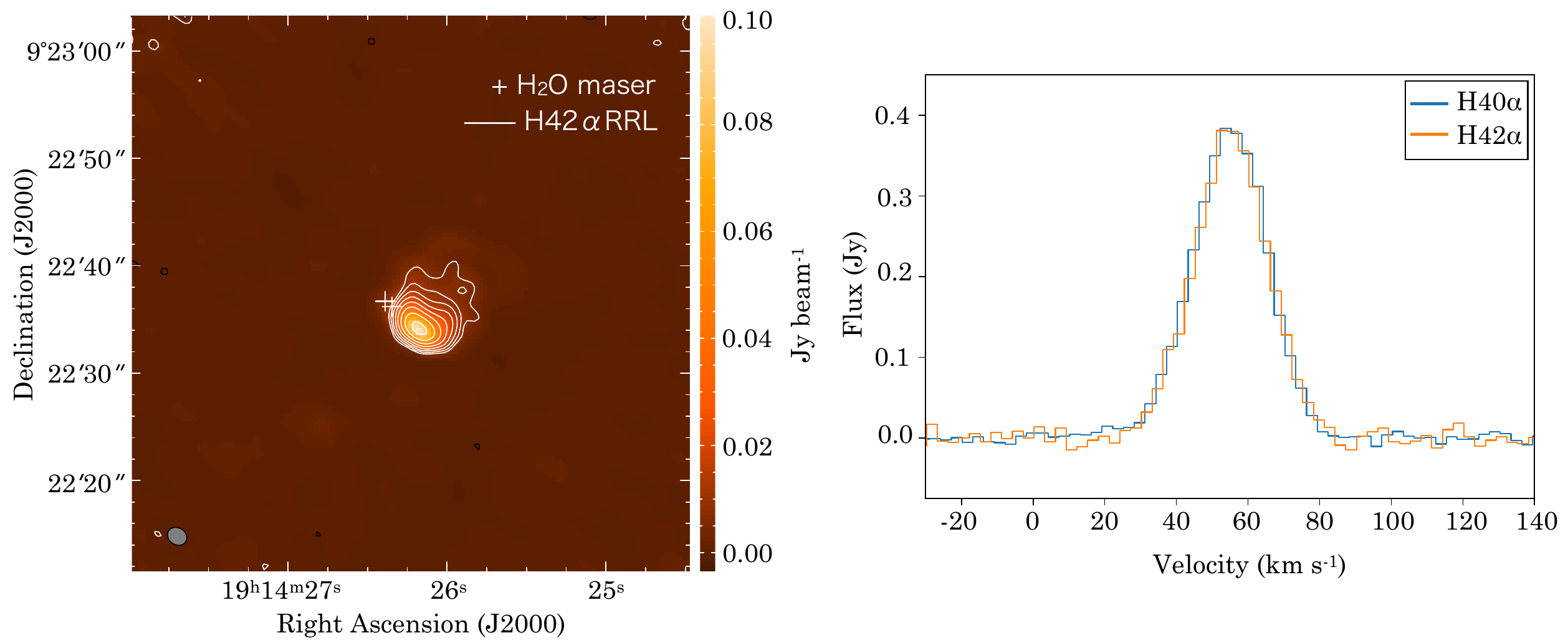}
    \caption{Continuum at $\nu$ $\sim$92.856~GHz and RRL emission of G43.89--0.78 obtained from ALMA observations. {\bf Left}: The colour map corresponds to the continuum emission at 0.3 cm and the contours represent the velocity--integrated emission of the H42$\alpha$ line obtained by integrating emission in the velocity range from $\sim$~0 to 100~km~s$^{-1}$, ensuring that all the line is included. The contours increase as $\pm$5$\times$rms$\times$1.5$^{i}$ with $i$=0,1,2,3$\ldots$ (black contours have negative values) and the rms is listed on Table~\ref{tab:tab2}. The synthesized beam (see Table~\ref{tab:tab2}) is shown in the bottom--left corner. The crosses mark the position of the water masers from \citet{Hofner1996}. {\bf Right}: Continuum subtracted line profiles of the H40$\alpha$ and H42$\alpha$ emission.}
    \label{fig:fig3}
\end{figure*}

Following \citet{Kurtz1999}, we see that for both the extended and the compact emission (the former traced by NVSS, the latter by the 0.7, 2 and 3.6 cm images), there is continuity at the peak--1 position (see Fig.~\ref{fig:fig1} and \ref{fig:fig2}).  Moreover, a result from \citetalias{delaFuente2020} is that 24~$\mu$m emission coincides with the RC emission at 3.6~cm at several scales --- including the saturated UC component.  This MIPS emission could indicate the presence of ionized gas at intermediate scales between the VLA high--resolution and low--resolution observations. VLA observations at 3.6~cm in the C or D-configuration could confirm this. MIPS emission at peak--1 is between 20\arcsec ~(the saturated zone) and 70\arcsec ~(including first Airy ring), in agreement with the size of 45$\arcsec$ (1.75~pc) obtained from the NVSS image.  The physical parameters of the EE and the \uchii are determined in the standard way (equations \ref{eqn:eqnA1} to \ref{eqn:eqnA4} in Appendix~\ref{sec:appendixA}), and these results are shown in Table~\ref{tab:tab3}. 

These physical parameters agree with those reported in Table~\ref{tab:tabA1} for sources in \citetalias{delaFuente2020} and \citet{Kim2001} as  shown in Appendix~\ref{sec:appendixA}. Considering the calculated values and the morphological study described above, G43.89--0.78 can be catalogued as an \uchiie. However, intermediate resolution VLA observations at 3.6~cm, such as those presented in \citetalias{delaFuente2020}, would be useful to strengthen this result.

\begin{figure*}
\centering
	\includegraphics[width=\textwidth]{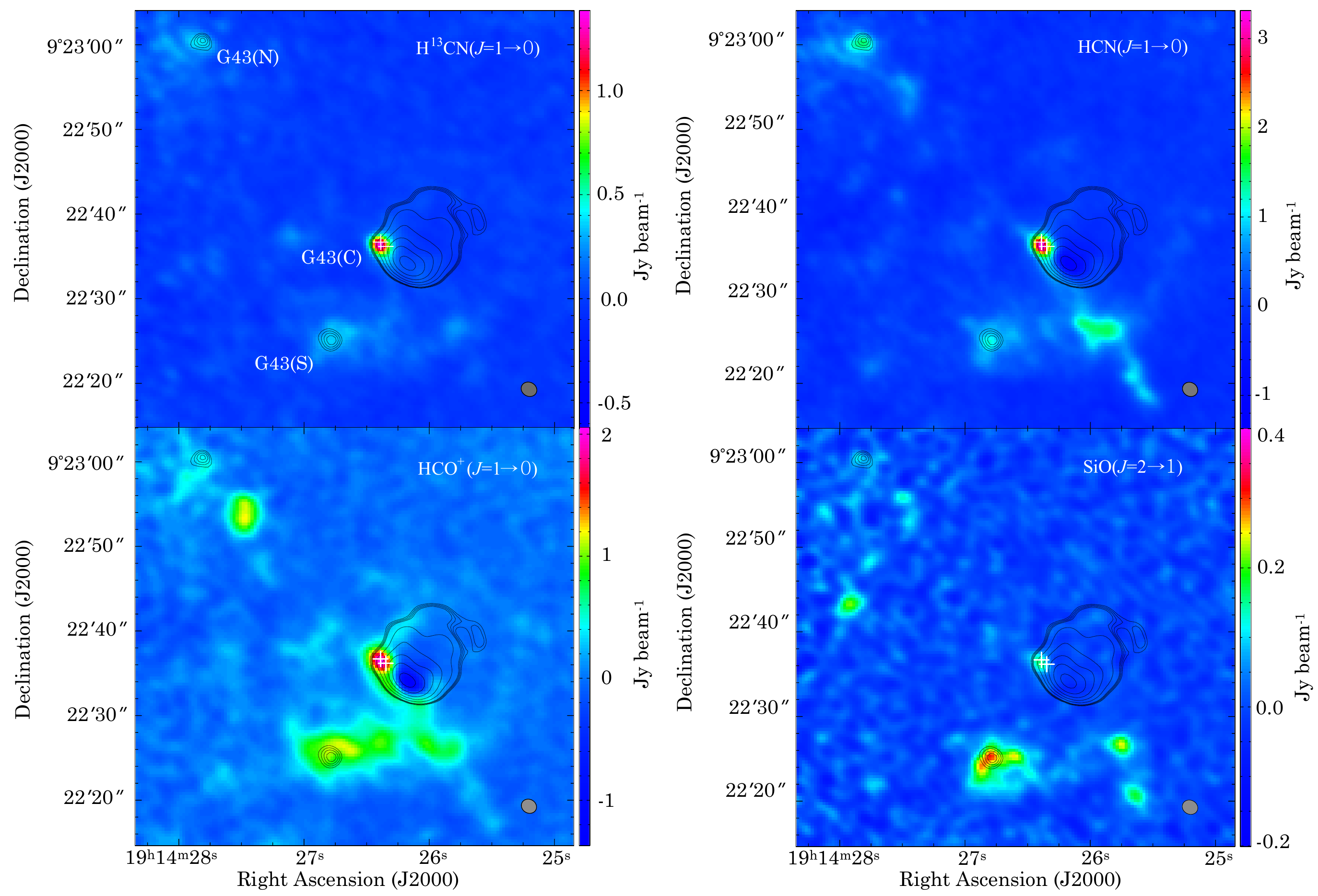}
    \caption{ALMA observations of the continuum emission as well as line emission of different molecular species in the vicinity of G43.89$-$0.78. The black contours represent the continuum emission at 0.3 cm. The first four contours are 5, 6, 7 and 8 times the rms of the map (listed in Table~\ref{tab:tab2}) and then they increase as $\pm$8$\times$rms$\times$2$^{i}$ with $i$=0,1,2,3$\ldots$. Apart from the cometary UC HII region, three other continuum clumps were identified and they are indicated in the upper--left panel as G43(N), G43(C) and G43(S) (see main text). The water masers from \citet{Hofner1996} are marked as white crosses.}
    \label{fig:fig4}
\end{figure*}

\begin{table*}
\caption{Physical parameters (continuum) for clumps on G43.89 vicinity}
\label{tab:tab5}
\begin{threeparttable}
\begin{tabular}{lcccccccc}
\hline

\, Clump\tnote{a} & Other\tnote{b} & Right Ascension & Declination & S$_{\nu}$\tnote{c} & Source\tnote{d} & T$_{\rm dust}$ & M$_{\rm H_2}$ & 24$\mu$m  \\

\, Designation & Designation & (J2000:\, h$\,\,$m$\,\,$s) & (J2000:\, $\circ \,\, \prime \,\, \prime\prime$) & (mJy) & size (\arcsec)  & (K)  & (M$_{\odot}$) & Counterpart     \\
\hline
\, G43.899--0.786 & G43 N & 19 14 27.818 & +09 23 00.33 &6.00 & 1.34 & 50.00 & 11.00 & Y    \\
\, G43.888--0.787 & G43 S & 19 14 26.789 & +09 22 25.12 &8.00 & 2.60 & 50.00 & 15.00 & N     \\
\hline
\end{tabular}
\begin{tablenotes}

\item[a] Based on galactic coordinates. \\
\item[b] Respect to the main clump: G43(C) or G43.890--0.784,  \\
\item[c] Integrated flux. Uncertainties of 5\%. \\
\item[d] Deconvolved size obtained from observations using CASA. $\Theta_{\rm s}$ = $\sqrt{\Theta_{\rm x} \Theta_{\rm y}}$   \\

\end{tablenotes}
\end{threeparttable}
\end{table*}

\begin{figure}
\centering
\includegraphics[width=\columnwidth]{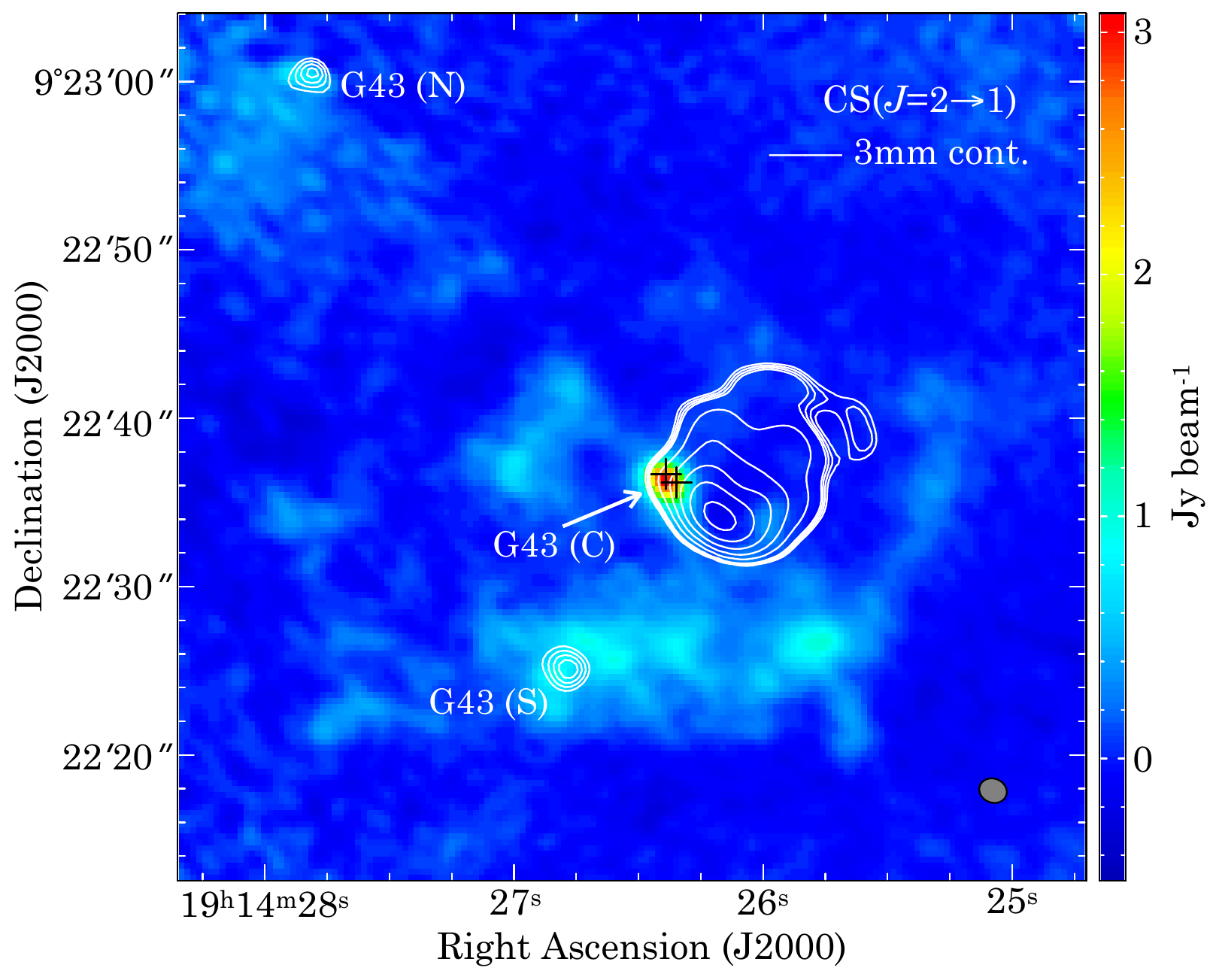}
    \caption{Continuum emission in G43.89$-$0.78 obtained from the ALMA observations. The colour map corresponds to the velocity--integrated CS($J$=2$\rightarrow$1) emission integrated over the velocity range 15$<$V(km~s$^{-1}$)$<$95. The rms of the image is 8.5$\times$10$^{-2}$~Jy~beam$^{-1}$ $\cdot$ km~s$^{-1}$. The contours represent the continuum emission at $\nu$ $\sim$92.856~GHz. The first four contours are 5, 6, 7 and 8 times the rms of the map (listed on Table~\ref{tab:tab2}) and then they increase as 8$\times$rms$\times$2$^{i}$ with $i$=0,1,2,3$\ldots$. The ellipse at the bottom--right corner represents the size of the synthesized beam given in Table~\ref{tab:tab2}. The positions of the water masers reported by \citet{Hofner1996} are indicated with black crosses and coincide with the molecular clump labelled as G43(C). Two extra radio--continuum clumps labelled as G43(N) and G43(S) are observed in the field.}
    \label{fig:fig5}
\end{figure}

\begin{figure*}
\centering
    \includegraphics[width=2\columnwidth]{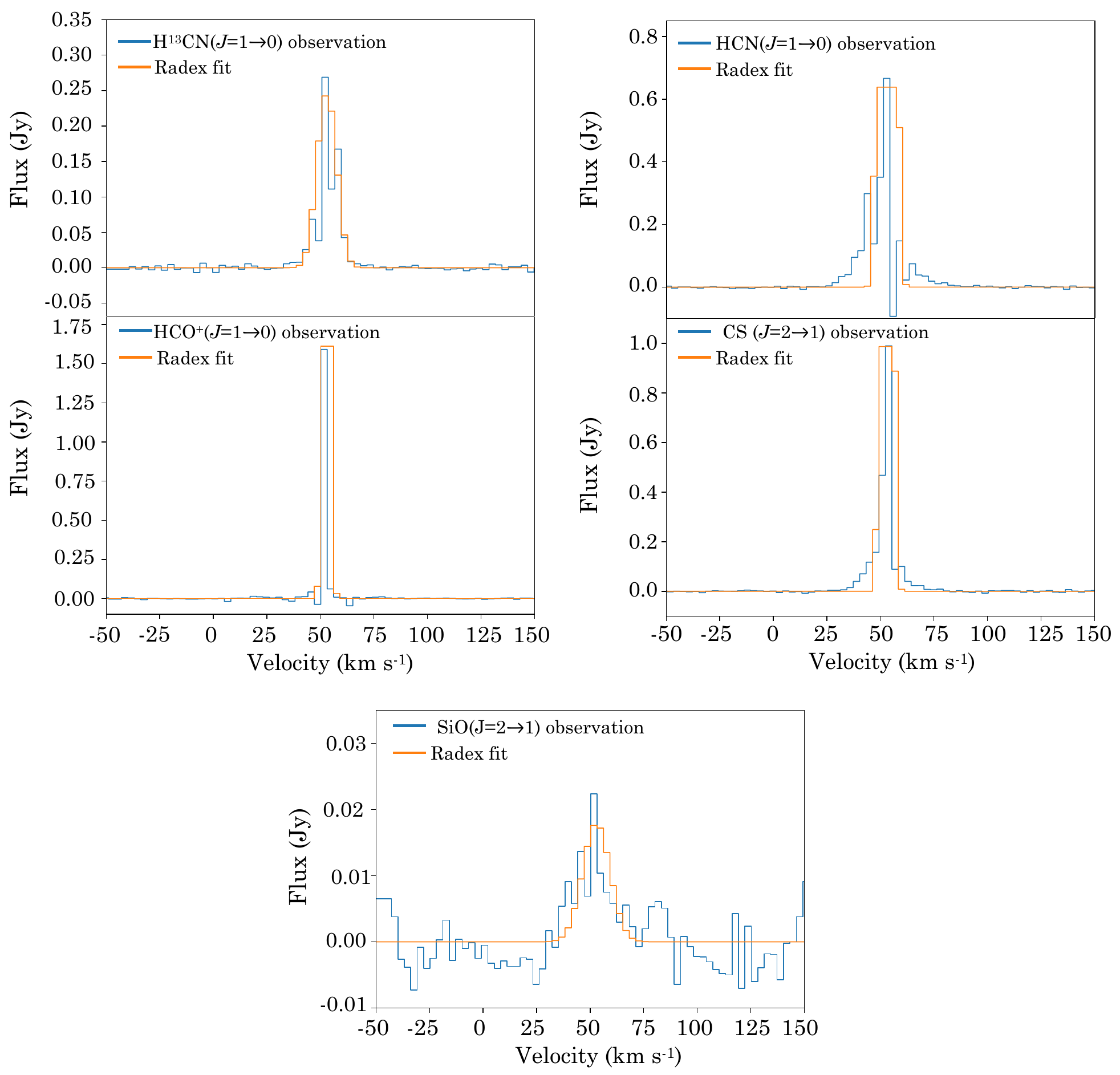}
    \caption{Spectral line profiles of different molecular species detected in G43.89(C) or G43.890--0.784. The blue lines are the ALMA observations and the orange lines are fits to the observations obtained using the RADEX in CASSIS (see Table \ref{fig:fig6} for the parameters of the fit models).}
    \label{fig:fig6}
\end{figure*}

\citet{Garcia1996} show that wind driven bubbles formed by the expansion of the \uchii regions produce structures that are not larger than a scale of $\sim$~2~pc. From Fig.~\ref{fig:fig1} it is clear that the RC emission of the \uchii in G43.89 extends beyond $\sim$~4~pc. Therefore, it is likely that the expansion of a bubble driven by a wind cannot be applied to explain all of the \textit{Spitzer} bubbling structures observed nearby G43.89. 

On the other hand, star forming regions with molecular nebulosity from the parent cloud(s) have been modelled as the result of the collision of two molecular clouds triggering massive star formation; i.e., the cloud--cloud collision (CCC) mechanism \citep[][]{Nagasawa1987,Habe1992}. This model has been applied to giant molecular clouds turning into  rich clusters of high--mass stars \citep[e.g.,][]{Furukawa2009,Ohama2010,Fukui2014,Baug2016,Kuwahara2020}, and to smaller molecular clouds that trigger the formation of clusters with one or a few (OB type) massive stars \citep[e.g.,][]{Torii2011,Torii2015,Torii2017,Fujita2019}. With high resolution CO velocity maps it would be viable to model emission at large scales ($\gtrsim$~2~pc) around \uchiies ~with the CCC scenario.

In several cases in \citetalias{delaFuente2020} we note the presence of molecular nebulosity, with the embedded UC component resembling the clump at a shocked interface between colliding clouds. The extended emission also resembles the bubble structures left behind where one or more newly-formed stars ionize a cavity inside the larger cloud ({\it Spitzer} bubbles). Such a study could confirm if the RC morphological classification of \uchiies ~is related with the clouds' initial conditions, such as mass, velocity dispersion, velocity directions and viewing angle \citep[see ][ for their ``cup" and ``tunnel" orientation discussion]{Torii2015}. In particular, for the case of G43.89, this model could explain the presence of peaks--1 and 2 in the same region.

\subsection{The Ultracompact vicinity}
\label{sec:sec3.2}

\subsubsection{RC and RRL emission}
\label{sec:sec3.2.1}

The high--resolution RC maps at 0.7, 2, and 3.6 cm, tracing the UC component, are presented in Fig.~\ref{fig:fig2}. These maps clearly show the cometary morphology of the \uchii region. The positions of the two water maser features reported by \cite{Hofner1996} are indicated with crosses on the images. The water masers are not spatially coincident with the peak of RC observed with the VLA, but rather are offset from the head of the cometary structure. Thus, the water masers are not coincident with the ionized gas. The fact that there is 3.6~cm RC emission partially overlapping with the water masers is due to the limited angular resolution of these observations. Indeed, convolving our higher angular resolution maps at 2 and 0.7~cm with a larger beam reproduces this effect. In addition, \cite{Hofner1996} mention that they found the masers to be ``well--separated'' from the cometary ionized arc, which is a result of their relatively high angular resolution ($\sim$ 0$\rlap{.}^{\prime\prime}$5) observations.

\begin{table*}
\centering
\caption{Results of the molecular line observations of G43(C) or G43.890--0.784.}
\label{tab:tab6}
\begin{threeparttable}

\begin{tabular}{lccrccc}
\hline

\,  Line & Peak\tnote{a} & V$_{\rm LSR}$ & $\Delta$V (FWHM) & Column Density & Abundance & Temperature\\
\,  &  (Jy) &  (km s$^{-1}$) & (km s$^{-1}$) & (cm$^{-2}$) & & (K) \\

\hline

\, $^{13}$CS($J$=2$\rightarrow$1) &  $<$0.035 & $\ldots$ & $\ldots$ &$\ldots$&$\ldots$ & $\ldots$\\
\,  CS($J$=2$\rightarrow$1) & 1.05 & 55.3 $\pm$0.1& 4.0 &1.00$\times$10$^{16}$&5$\times$10$^{-9}$ & 18\\
\,  HCN($J$=1$\rightarrow$0) &  0.68 &55.3 $\pm$0.1& 5.0 &1.00$\times$10$^{16}$&5$\times$10$^{-9}$ & 16\\
\,  H$^{13}$CN($J$=1$\rightarrow$0) & 0.27 & 55.3 $\pm$0.1& 10.0&1.67$\times$10$^{14}$&8$\times$10$^{-11}$ & 16\\
\,  HCO$^{+}$($J$=1$\rightarrow$0) & 1.60 & 55.3 $\pm$0.1& 3.0&4.00$\times$10$^{15}$&2$\times$10$^{-9}$ & 35\\
\,  SiO($J$=2$\rightarrow$1) &  0.025 &  55.3 $\pm$0.1& 15.0 &1.60$\times$10$^{13}$&8$\times$10$^{-12}$ & 18\\
\hline

\end{tabular}
\begin{tablenotes}
\item[a] Uncertainties on integrated flux are 20\%. Typical OVRO beam sizes for the 0.3 cm lines are 3\arcsec. Uncertainties on integrated flux for ALMA observations are 5\%. \\
\item[b] Deconvolved size obtained from observations using task \textit{imfit} of AIPS. $\Theta_{\rm s}$ = $\sqrt{\Theta_{\rm x} \Theta_{\rm y}}$  \\

\end{tablenotes}
\end{threeparttable}
\end{table*}

Fig.~\ref{fig:fig3} shows the 0.3~cm continuum, H40$\alpha$, and H42$\alpha$ RRL emission obtained from ALMA observations. A contour map of the velocity--integrated H42$\alpha$ RRL emission, superimposed on a colour map of the continuum emission, is shown in the left panel, while the profiles of the H40$\alpha$ and H42$\alpha$ lines are shown in the right panel. The velocity--integrated map of the H42$\alpha$ RRL was obtained by integrating emission in the velocity range from $\sim$0 to 100~km~s$^{\rm -1}$. A Gaussian fit to the profiles of the H40$\alpha$ and H42$\alpha$ RRLs gives a peak velocity V$_{\rm LSR}$~=~55.2$\pm$0.1~km~s$^{-1}$ and a FWHM $\Delta$V~=~24.3$\pm$0.3~km~s$^{-1}$, which is in agreement with the systemic velocity V~$\sim$~55~km~s$^{-1}$ obtained from previous molecular observations \citep{Olmi1993,Bronfman1996,Olmi1999,Araya2002}. A fit to the H41$\alpha$ line observed with OVRO gives a similar result (see Table~\ref{tab:tab4}). These results indicate that the ionized gas and the molecular gas are physically associated.

From the ratio of the continuum and RRL fluxes, and the line--width of the RRL, it is possible to determine the LTE electronic temperature, ${\rm T}_{\rm e}$, of the ionized gas as \citep{Mezger1967,Dupree1970}:

\begin{equation}
\label{eqn:eqn1}
\left(\frac{\mathrm{T_e}}{\mathrm{K}}\right) = 2.2\times 10^3 \left( \frac{\nu}{\mathrm{GHz}}\right)^{0.96}
\left(\mathrm{\frac{T_L}{T_C}}\right)^{-0.87}
\left(\frac{\Delta \mathrm{V}}{\mathrm{km\,s^{-1}}}\right)^{-0.87},
\end{equation}

\noindent  where $\nu$ is the observation frequency, T$_{\rm C}$ is the brightness temperature of the continuum emission, and T$_{\rm L}$ and $\Delta$V are the continuum-subtracted peak brightness temperature and the line--width of the RRL, respectively. Since S$_{\rm L}$/ S$_{\rm C}$ is $\sim$0.95 (see Table~\ref{tab:tab4}), equation \ref{eqn:eqn1} gives an LTE T$_{\rm e}$~$\sim$~1.1$\times$10$^4$~K. The physical parameters obtained for the \uchiir in G43.89, using this temperature value, are presented in Table~\ref{tab:tab3}. For the calculations we used the 3.6~cm data and a deconvolved size of the source $\simeq$~3$^{\prime\prime}$ ($\simeq$~0.12~pc). The derived values of the physical parameters for this source are in approximate agreement with those obtained by \citet{Wood1989}. 

\begin{figure*}
\centering
	\includegraphics[width=\textwidth]{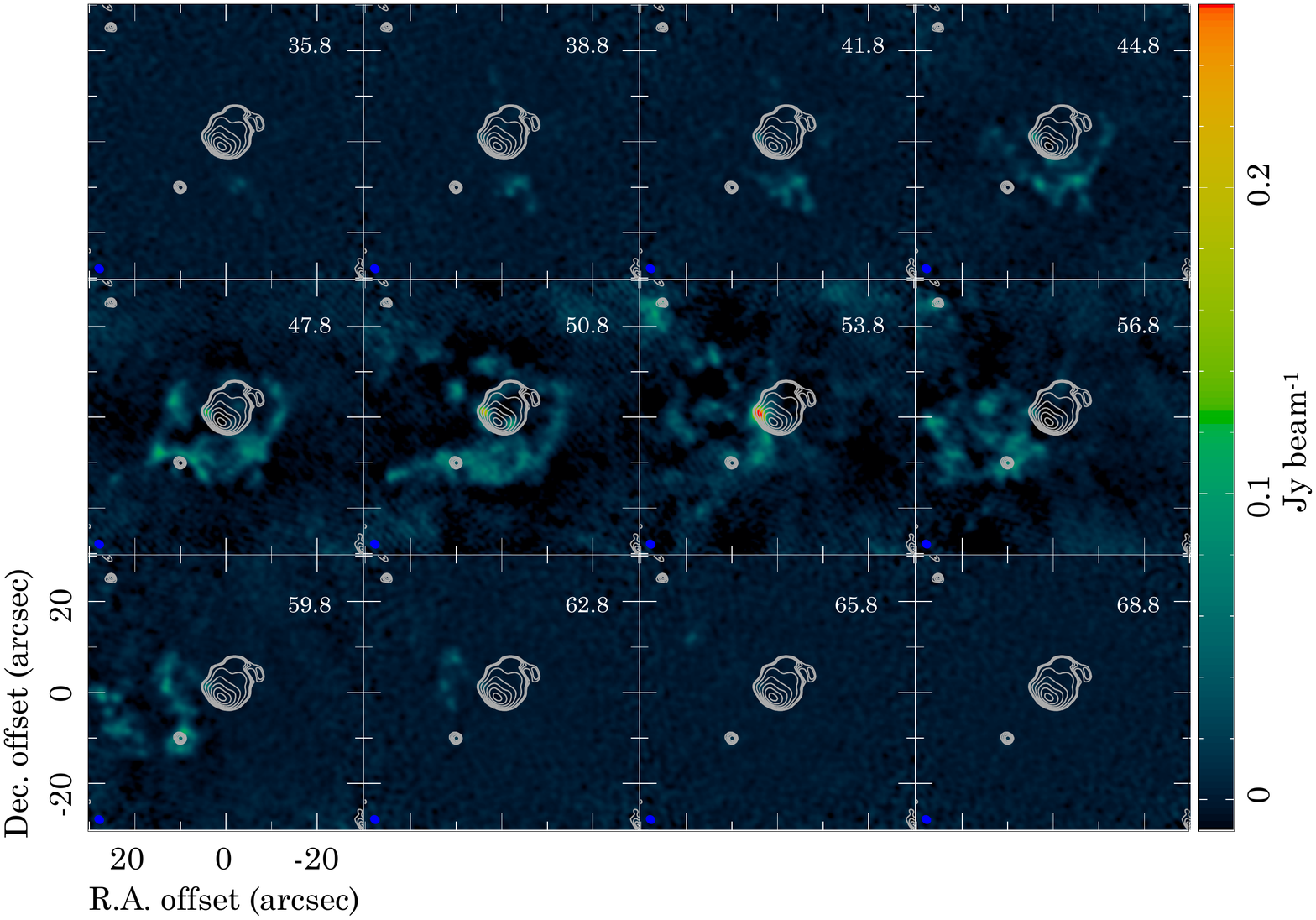}

    \caption{Channel maps of the ALMA CS emission in G43.89$-$0.78. The colour-map is the CS($J$=2$\rightarrow$1) line emission in the vicinity of G43.89$-$0.78 for individual channel maps. The number in the top-right corner is the velocity of the channel. The white contours represent the continuum emission at 3mm. The first four contours are 5, 6, 7 and 8 times the rms of the map (listed on Table~\ref{tab:tab2}) and then they increase as $\pm$8$\times$rms$\times$2$^{i}$ with $i$=0,1,2,3$\ldots$. The synthesised beam is indicated with a blue ellipse located in the bottom-left corner of the individual channel maps (see Table \ref{tab:tab2}).}
    \label{fig:fig7}
\end{figure*}

\begin{figure*}
\centering
	\includegraphics[width=\textwidth]{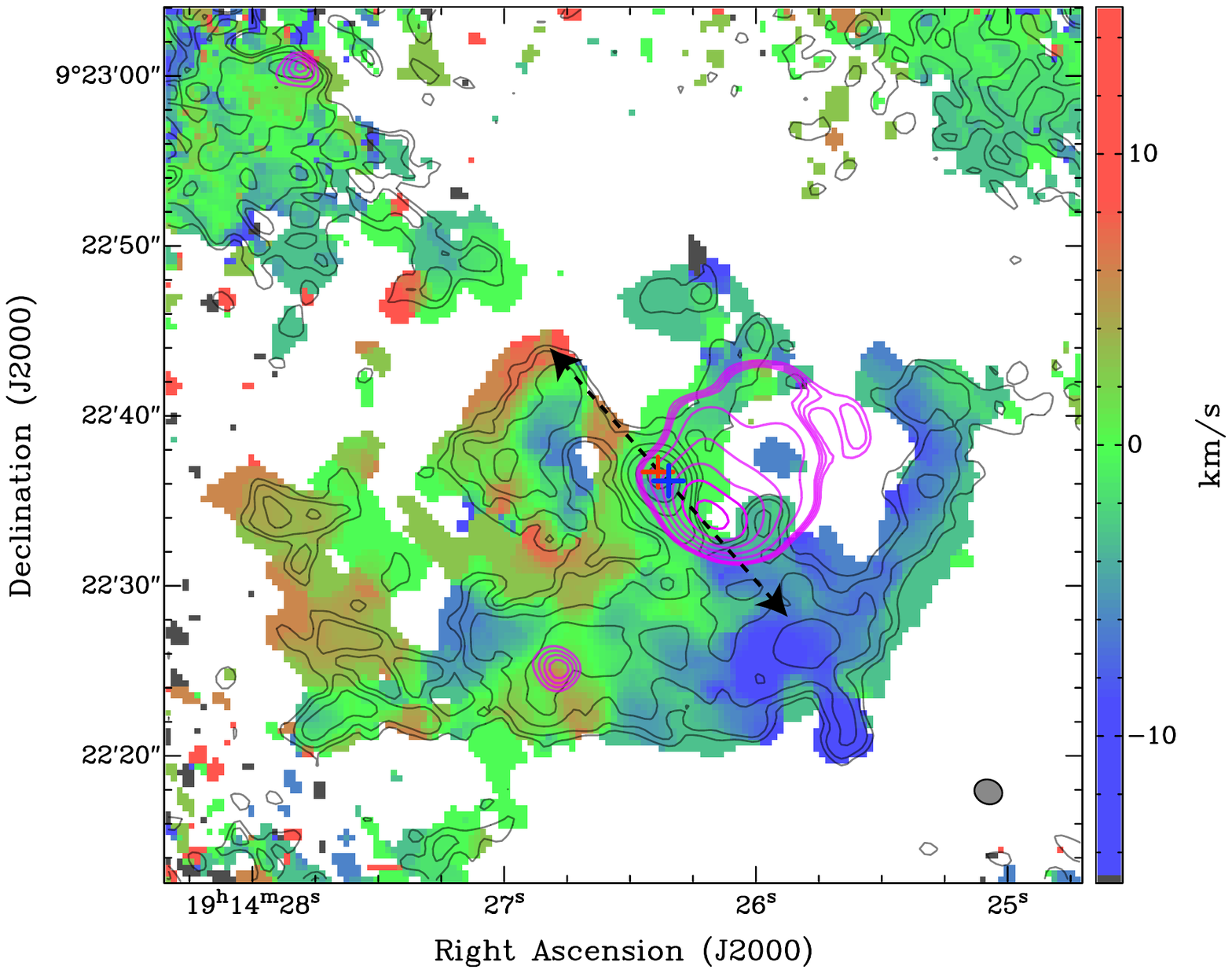}
    \caption{Moment-1 image of the  CS($J$=2$\rightarrow$1) line emission in the vicinity of G43.89$-$0.78. The colour-map is the moment-1 (velocity field) clipped at 0.01~Jy~beam$^{-1}$ (4 times the rms noise level of the channel maps). The black contours indicate the velocity-integrated CS($J$=2$\rightarrow$1) emission (moment-0). The contours increase as $\pm$5$\times$rms$\times$1.5$^{i}$ with $i$=0,1,2,3,$\ldots$, where the rms is 0.075~Jy~beam$^{-1}$~km~s$^{-1}$. The magenta contours represent the continuum emission at 3mm. The first four contours are 5, 6, 7 and 8 times the rms of the map (listed on Table~\ref{tab:tab2}) and then they increase as $\pm$8$\times$rms$\times$2$^{i}$ with $i$=0,1,2,3$\ldots$. A broken line with arrows indicates the direction of a kinematical feature associated with the clump G43(C) that could be a molecular outflow (see main text).} 
  
    \label{fig:fig8}
\end{figure*}

\subsubsection{ALMA and OVRO: Molecular and mm--continuum emission}
\label{sec:sec3.2.2}

Using the JCMT, \citet{Hatchell1998} searched for molecular emission toward 14 \uchiirs including G43.89. For G43.89 they detected several molecular lines (C$^{17}$O, C$^{18}$O, SO, C$^{34}$S), and weak emission from the lowest excitation energy lines of CH$_{3}$CCH and CH$_{3}$OH. Also, \citet{Li2015} searched for sulphur--bearing molecules, including OCS, CS, H$_{2}$S and SO; they only detected emission from SO(5$_{6}$-4$_{5}$). 

From the ALMA observations, emission from CS($J$=2$\rightarrow$1), HCN($J$=1$\rightarrow$0), H$^{13}$CN($J$=1$\rightarrow$0), HCO$^{+}$($J$=1$\rightarrow$0), SiO($J$=2$\rightarrow$1) lines were detected (see Figs.~\ref{fig:fig4} and \ref{fig:fig6}). On the other hand, no OVRO $^{13}$CS($J$=2$\rightarrow$1) line emission was detected above a level of 3$\sigma$ ($\sigma$~=~0.35~mJy~beam$^{-1}$) toward the \uchii region. This is almost certainly due to the limited sensitivity of OVRO observations with respect to ALMA.

Aside from the \uchii region, two more radio--continuum clumps were detected in the field of view of the ALMA observations. They are marked as G43(N) and G43(S) in Figs.~\ref{fig:fig4} and \ref{fig:fig5}. By their galactic coordinates, these sources are designated as G43.899--0.786 and G43.888--0.787, respectively. G43(N) is coincident with a 24$\mu$m point--source within the positional uncertainties.  This clump could be a proto--stellar source.  On the other hand, G43(S) does not have a 24$\mu$m counterpart, although there is considerable molecular emission at its position.  G43(N), in this regard, shows only weak molecular emission. Physical parameters for these clumps are reported in Table~\ref{tab:tab5}.

In addition, close inspection of the continuum emission associated with the \uchii region reveals that there is a bump toward the position of the water masers, which also coincides with a bright molecular clump. We label this source as G43(C) with galactic designation G43.890--0.784. 
Figs.~\ref{fig:fig4} and \ref{fig:fig5} shows that except for the SiO($J$=2$\rightarrow$1) emission, the molecular emission peaks in G43(C). In these figures, it is clear that the water masers are associated with G43(C). The coordinates of this brightest clump are R.A.(J2000)~=~19h 14m 26.387s, Dec.(J2000)~=~9$^{\circ}$ 22$^{\prime}$ 36$\rlap{.}^{\prime\prime}$28. The configuration of the cometary \uchiir with a molecular clump and water masers resembles that of G12.21, where \citet{delaFuente2018} concluded it is a hot molecular core. Thus, in order to assess whether G43.89 also contains a hot molecular core we proceed to study the molecular clump of G43.89 in a similar way as for G12.21.

To determine the physical parameters of the molecular gas in G43(C) we used the CASSIS software, the CDMS spectroscopic databases, the LAMDA molecular databases, and the RADEX code\footnote{CASSIS has been developed by IRAP-UPS/CNRS (http://cassis.irap.omp.eu)}. We extracted spectra from all the molecular species detected with ALMA and then we fitted a model by varying parameters such as the kinetic temperature, and column densities. Fig.~\ref{fig:fig6} shows the fits to the observations. The best fit gives a molecular hydrogen column density of N(H$_2$) = 2$\times$10$^{24}$ cm$^{-2}$, which translates into a volumetric density of n(H$_2$) = 2.6$\times$10$^{7}$~cm$^{-3}$ and a mass of M(H$_2$) = 220~M$_{\odot}$. The physical parameters of the best fit models are listed in Table~\ref{tab:tab6}. The kinetic temperature of the gas ranges from 16 to 35~K. The gas traced by HCO$^{+}$ is the warmest of all, suggesting that the emission arises closer to a proto--stellar object. 

The G43(C) clump does not completely meet the operational definition of a hot molecular core given in  \citet[][and references therein]{Garay1999}: size~$\lesssim$~0.1~pc, T$_{\rm k}$~>~50~K, and n$_{\rm e}\gtrsim$~10$^5$~cm$^{-3}$, because the reported temperature is $\lesssim$~50~K. Possibly this YSO is in such an early evolutionary stage that the gas has not yet been heated by the YSO or that the dusty molecular core lacks sufficient optical depth to absorb the radiant energy. In an attempt to estimate the evolutionary stage of G43(C), we compare the physical parameters from Table~\ref{tab:tab6} with the chemical models from \citet[][]{Nomura2004} assuming a temperature-dependent grain mantle evaporation process. An approximate timescale of 1$\times$10$^4$~yr is obtained for a column density of CS 1$\times$10$^{16}$~cm$^{-2}$.

A timescale of $\sim$~6$\times$10$^4$~yr is obtained from HCN (see their Fig.~2(c)). In comparison, the average column densities of some molecular tracers from hot cores given by \citet[][see their Table~6]{Li2015} are more consistent with a CS timescale between 6$\times$10$^4$ and 1$\times$10$^5$~yr, which corresponds to an older stage than for G43(C) by about an order of magnitude. This is in agreement with the chemical evolutionary sequence proposed by \citet{Beuther2007} for hot molecular cores, where cores at $\sim$~30~K cause some molecules to be released from dust grains (such as CS), but when the temperature of the hot core increases to $\gtrsim$~100~K, the H$_2$O molecules released also from dust grains generate the enrichment of other molecular tracers and their abundance (e.g., SO). 

We also estimated the mass of the clumps G43(N) and G43(S). For a frequency of 86~GHz, considering continuum flux densities (S$_{\nu}$), and assuming emission from isothermal, optically thin dust, the gas mass can be determined using  \citep[][]{Hildebrand1983,Gall2011}:

\begin{equation}
\label{eqn:eqn2}
        \Bigg[\frac{M_{\rm gas}}{M_{\sun}}\Bigg] = 4.8 \times 10^{-23} \frac{g}{B_{\nu}(T_{\rm dust})}\Bigg[\frac{S_\nu^{\rm N,S}}{\rm mJy}\Bigg] \Bigg[\frac{\kappa_\nu}{\rm cm^{2}g^{-1}}\Bigg]^{-1} \Bigg[\frac{D}{\rm pc}\Bigg]^2 ,
\end{equation}
where $g$ is the gas--to--dust ratio, $D$ the adopted distance, $\kappa_{\nu}^{1.3mm}$ the dust mass opacity coefficient, and $B_\nu(T_{\rm dust}$) the Planck function for a dust temperature $T_{\rm dust}$. For a $B_\nu(T_{\rm dust}$) with T$_d$~=~50~K, we derive the clump masses reported in Table~\ref{tab:tab5}. In this calculation, the uncertainty is only referenced to the uncertainty in the adopted value of $\kappa_{\nu}^{1.3mm}$; we use 0.015 cm$^{2}$ gr$^{-1}$. 

\subsubsection{A molecular outflow from the clump G43(C)?}
\label{sec:sec3.2.3}

From the line profiles shown in Fig. \ref{fig:fig6} it can be seen that some molecular species, such as HCN and CS, exhibit wings with relatively high velocities (spanning a velocity range of $\sim$50~km~s$^{-1}$). Such wings cannot be fitted with any Gaussian curve. Furthermore, the CS($J$=2$\rightarrow$1) emission in the vicinity of G43.89$-$0.78 exhibits a NE-SW velocity gradient, which can be seen in Figs. \ref{fig:fig7} and \ref{fig:fig8}.

In particular, there is a feature that extends in the NE-SW direction that connects G43(N) with G43(C) and a clump of molecular emission located SW of G43(C). In addition, the velocity gradient of the water masers associated with G43(C) (see Fig. \ref{fig:fig8}) matches the velocity gradient of the molecular emission. This suggests that G43(C) hosts a bipolar molecular outflow that may be the origin of the water masers, as has been observed in other sources \citep[][]{Trinidad2009}. The direction of the outflow is indicated with a broken line in Fig. \ref{fig:fig8}. Further investigations with higher angular resolution, involving proper motion observations, are needed to confirm this idea and to elucidate the nature of this object. 

\section{Summary and Conclusions}
\label{sec:summary}

\begin{enumerate}

\item The EE associated with G43.89 meets the nominal definition of  \uchiies based on physical parameters: EM $\sim$~10$^{\rm 4}$--10$^{\rm 5}$~cm$^{\rm -6}$~pc, n$_{\rm e}$~$\lesssim$~5$\times$10$^{2}$~cm$^{\rm -3}$, and M$_{\rm HII}$ $\sim$ 5 to 10$^3$~M$_\odot$, therefore it can be added to the final catalogue given in \citetalias{delaFuente2020}.

\item G43.89--0.78 is confirmed as an \uchiir using high--resolution VLA data at 3.6 cm. It has a dense (2.6$\times$10$^7$ cm$^{-3}$), and small ($\sim$2'' or 0.08 pc) nearby molecular clump with mass of 220~M$_{\odot}$. The water masers appear to be spatially coincident with this clump, and may arise in a molecular outflow.

\item The low temperature of this clump implies that it is not a hot molecular core, however a timescale of few times 10$^4$~yr is estimated by comparison with chemical models. This value is smaller than the average timescale for typical hot cores by about an order of magnitude.

\item G43.890--0.784 is rich in molecular species, and we detect two new radio--continuum clumps, G43.899--0.786 and G43.888--0.787 respectively, located in its vicinity.

\end{enumerate}

\section*{Acknowledgements}

The National Radio Astronomy Observatory is a facility of the National Science Foundation operated under cooperative agreement by Associated Universities, Inc. This work is based in part on observations made with the {\it Spitzer} Space Telescope, which is operated by the Jet Propulsion Laboratory, California Institute of Technology under a contract with NASA. ALMA is a partnership of ESO (representing its member states), NSF (USA), and NINS (Japan), together with NRC (Canada), MOST, and ASIAA (Taiwan), and KASI (Republic of Korea), in cooperation with the Republic of Chile. The Joint ALMA Observatory is operated by ESO, AUI/NRAO, and NAOJ.  EdlF acknowledges support from Consejo Nacional de Ciencia y Tecnolog\'ia (CONACyT), Mexico: grant 124449, and CONACyT--SNI exp. 1326. EdelF also thanks Chalmers University of Technology, Onsala Space, Sweden for the support during several research stays. M.A.T. acknowledges support from Universidad de Guanajuato grant DAIP--33/2019. AP acknowledges INAOE/CONACyT financial support for authorship while home--working during the COVID--19 pandemic.

\section*{Data availability}

The data underlying this article are available in:

1.- ALMA archive at https://almascience.eso.org/asax/, and can be accessed with ALMA\#2015.1.00280.S.

2.- NRAO Science Data Archive [Karl Jansky VLA Database] at https://archive.nrao.edu/archive/archiveproject.jsp, and can be accessed with the respective Project (Proposal) Code: AK423

3.- IRSA NASA/IPAC Infrared Science Archive at https://irsa.ipac.caltech.edu/Missions/spitzer.html, and can be accessed by selection of GLIMPSE II (program ID=187), Data Set Identification = ads/sa.spitzer\#0011972096; and  MIPSGAL, program IDs 20597 and 30592.

4.- OVRO data will be shared on reasonable request to the corresponding author.





\appendix
\section{Physical parameters (3.6 cm) of \uchii regions with extended emission}
\label{sec:appendixA}

\begin{table*}
\caption{Physical Parameters of the \uchiies reported in Table 7 of \citetalias{delaFuente2020}, using RC observations at 3.6 cm \tnote{a}$^{\rm a}$.}
\label{tab:tabA1}
\begin{threeparttable}
\begin{tabular}{lcccccr}
\hline

\, Source & S$_{\nu}$\tnote{a} & Source\tnote{b} & Source\tnote{b} & EM & n$_{\rm e}$ & M$_{\rm \hii}$  \\

\, Name & (Jy) & Size (\arcmin) & Size (pc) & (10$^4$cm$^{-6}$ pc) & (cm$^{-3}$) & (M$_{\odot}$)     \\

\hline

\, G05.48--0.24  & 1.18 &  2.45 & 3.10 & 2.50 & 40.80 & 1799.55   \\
\, G05.97--1.17  & 8.06 &  2.65 & 2.08 & 14.80 & 218.40 & 81.96   \\
\, G10.30--0.15  & 5.80 &  2.45 & 4.28 & 12.50 & 139.80 & 454.96   \\
\, G12.21--0.10  & 1.30 &  3.46 & 13.59 & 1.40 & 26.30 & 2745.07  \\
\, G18.15--0.28  & 4.20 &  2.83 & 3.46 & 6.76 & 114.50 & 197.04   \\
\, G19.60--0.23  & 4.33 &  1.26 & 1.28 & 35.18 & 428.80 & 37.68   \\
\, G23.44--0.21  & 1.16  & 2.00 & 5.24 & 3.74 & 69.20 & 413.54  \\
\, G23.71+0.17   & 1.49 &  2.00 & 5.18  & 4.80 & 78.90 & 455.77   \\
\, G25.71+0.04 &  0.68  &  3.46 & 9.36 & 0.73 & 22.90 & 782.00    \\
\, G28.20--0.05 & 0.48 &  1.50 & 3.97 & 2.75 & 68.20 & 177.62   \\
\, G31.39--0.25 & 0.65 &  1.94 & 5.02 & 2.23 & 54.50 & 287.59   \\
\, G35.20--1.74 & 11.31 &  2.00 & 1.92 & 3.65 & 112.80 & 33.24  \\
\, G37.55--0.11 & 0.93 &  2.00 & 5.76  & 3.00 & 59.10 & 469.90  \\
\, G37.87--0.40  & 4.11 &  1.00 & 2,71 & 53.00 & 362.50 & 298.72   \\
\, G45.07+0.13 &  0.73 &  1.00 & 1.75  & 9.42 & 190.20 & 42.09   \\
\, G45.12+0.13 &  1.80 &  1.50 & 2.62 & 10.32 & 162.60 & 121.42   \\
\, G45.45+0.06 &  4.50 & 2.00 & 3.49  & 14.51 & 166.90 & 295.57   \\
\, G54.10--0.06 & 0.52 &  2.45 & 5.63 & 1.12 & 36.50 & 270.99    \\
\, G60.88--0.13 & 0.49 & 1.73 & 1.11 & 2.11 & 113.10 & 6.39  \\
\, G77.96--0.01 & 0.95  & 2.45 & 3.14 & 2.04 & 66.10 & 84.80   \\
\, G111.28--0.66 & 0.30 & 2.50 & 1.82 & 0.62 & 47.80 & 11.95   \\

\hline
\end{tabular}
\begin{tablenotes}
\item[a] We use the VLA conf. D data for all sources except G60.88--0.13 (conf. C). See \citetalias{delaFuente2020} for details. In this computing, we assume the canonical value for electron temperatures of 1 $\times$ 10$^4$ K for all sources. Distances are reported on Table 1 of \citetalias{delaFuente2020}. The ionizing photon rates and spectral types are shown {\bf in} Table 7 of \citetalias{delaFuente2020}. \\
\item[b] Size = $\sqrt{\Theta_x\Theta_y}$. The values of $\Theta_x$ and $\Theta_y$ were obtained from Table 5 of \citetalias{delaFuente2020}.   \\

\end{tablenotes}
\end{threeparttable}
\end{table*}

To obtain the physical parameters of the ionized gas, we assume an isothermal, pure hydrogen, homogeneous, spherically symmetric (radius $r$ and size $\Theta_{\rm S}$), optically thin \hii~ region, with a distance $D$ from the Sun and electronic
temperature T$_{\rm e}$. The emission measure (EM), electronic density (n$_{\rm e}$), the mass of the ionized gas (M$_{\rm HII}$), and the total rate of emission of  Lyman continuum photons of the ionizing star (N'$_{\rm c}$) were calculated in the standard way using equations \ref{eqn:eqnA1} to \ref{eqn:eqnA4} \citep[][]{Kurtz1994,Panagia1978,Schraml1969,Oster1961}. The spectral type was obtained using \citet{Panagia1973}.

The physical parameters of \uchiies from \citetalias{delaFuente2020} are shown in Table~\ref{tab:tabA1}. For them, the total rate of emission of Lyman continuum photons and ionizing star type are shown in \citetalias{delaFuente2020}. For these \uchiies, values are: size $\sim$ 1--4\arcmin    ($\sim$ 1--14 pc), EM $\sim$ 10$^{\rm 4}$--10$^{\rm 5}$ cm$^{\rm -6}$ pc, n$_{\rm e}$ $\lesssim$ 5$\times$10$^{2}$ cm$^{\rm -3}$, and M$_{\rm HII}$ $\lesssim$ 10$^{3}$ M$_\odot$. The values reported by \citet{Kim2001} at 20 cm are: size $\sim$ 1--13\arcmin  ($\sim$ 3--19 pc), EM $\sim$ 10$^{\rm 4}$ cm$^{\rm -6}$ pc, n$_{\rm e}$ $\lesssim$ 10$^{2}$ cm$^{\rm -3}$, and M$_{\rm HII}$ $\sim$ 10$^{2}$ to 10$^{3}$ M$_\odot$. Combining both samples, we confirm \uchiies are \hiirs with these values: sizes $\sim$ 1--10\arcmin ($\sim$ 1--20 pc),~ EM $\sim$ 10$^{\rm 4}$--10$^{\rm 5}$ cm$^{\rm -6}$ pc, n$_{\rm e}$ $\lesssim$ 5$\times$10$^{2}$ cm$^{\rm -3}$, and M$_{\rm HII}$ $\sim$ 5 to 10$^3$ M$_\odot$. 

Table 1 of \citet{Kurtz2002} summarizes the physical parameters of several types of \hiirs. Updating with information from  the 3.6 cm observations, we show an adaptation of this table, including \uchiies in Table~\ref{tab:tabA2}.

\begin{multline}
\label{eqn:eqnA1}
\left(\frac{\mathrm{n_e}}{\mathrm{cm^{-3}}}\right) = 7.8\times 10^3 \left( \frac{\nu}{\mathrm{4.9\, GHz}}\right)^{0.05}
\left(\mathrm{\frac{S_\nu}{mJy}}\right)^{0.5}
\left(\frac{\mathrm{T_e}}{\mathrm{10^{4}\, K}}\right)^{0.175}\\
\times \left(\mathrm{\frac{\rm {\Theta_s}}{arcsec}}\right)^{-1.5}
\left(\mathrm{\frac{D}{kpc}}\right)^{-0.5},
\end{multline}

\begin{multline}
\label{eqn:eqnA2}
\left(\frac{\mathrm{EM}}{\mathrm{cm^{-3}}}\right) = 4.4\times 10^5 
\left( \frac{\nu}{\mathrm{4.9\, GHz}}\right)^{0.1}
\left(\mathrm{\frac{S_\nu}{mJy}}\right)
\left(\frac{\mathrm{T_e}}{\mathrm{10^{4}\, K}}\right)^{0.35} \\
\times \left(\mathrm{\frac{\rm {\Theta_s}}{arcsec}}\right)^{-2},
\end{multline}

 \begin{multline}
\label{eqn:eqnA3}
\left(\frac{\mathrm{M_{\rm HII}}}{\mathrm{M_{\odot}}}\right) = 3.7\times 10^{-5}
 \left( \frac{\rm \nu}{\mathrm{4.9\, GHz}}\right)^{0.05}
\left(\mathrm{\frac{S_\nu}{mJy}}\right)^{0.5}
\left(\frac{\mathrm{T_e}}{\mathrm{10^{4}\, K}}\right)^{0.175}\\
\times \left(\mathrm{\frac{\rm {\Theta_s}}{arcsec}}\right)^{1.5}
\left(\mathrm{\frac{D}{kpc}}\right)^{2.5},
\end{multline}

\begin{equation}
\label{eqn:eqnA4}
\left(\frac{\rm N'_c}{\rm s^{-1}}\right)~\ge~8.04\times10^{46}\left(\frac{\rm T_e}{\rm K}\right)^{-0.85}\left(\frac{\rm r}{\rm pc}\right)^3\left(\frac{\rm n_e}{\rm cm^{-3}}\right)^2.
\end{equation}

\begin{table*}
\centering
\caption{Physical parameters of \hiirs from radio observations. The ultracompact \hii region with extended emission (\uchiie) category is added for the first time using 3.6 cm data.}
\label{tab:tabA2}
\centering
\begin{threeparttable}
\begin{tabular}{lccccr}
\hline

\, Class of \hii & Size & EM & n$_{\rm e}$ & M$_{\rm \hii}$  \\
\, region &  (pc) & (cm$^{-6}$ pc) & (cm$^{-3}$) & M$_{\odot}$  \\
\hline

\, Hypercompact & $\sim$ 0.003 & $\gtrsim$ 10$^{10}$  & $\gtrsim$ 10$^6$ &  $\sim$ 10$^{-3}$    \\
\, Ultracompact & $\lesssim$ 0.1  & $\gtrsim$ 10$^{7}$  & $\gtrsim$ 10$^{4}$  & $\sim$ 10$^{-3}$    \\
\, Compact & $\lesssim$ 0.5  & $\gtrsim$ 10$^{7}$  & $\gtrsim$ 5$\times$10$^{3}$  & $\sim$ 1    \\
\, \uchiie & 1--20  & 10$^4$--10$^5$ & $\lesssim$ 5$\times$10$^2$ & 5--10$^3$    \\
\, Classical & $\sim$ 10  & $\sim$ 10$^2$  & $\sim$ 100  & $\sim$ 10$^5$    \\
\, Giant & $\sim$ 100  & $\sim$ 5 $\times$ 10$^5$  & $\sim$ 10  & 10$^3$--10$^6$    \\
\, Supergiant & $\gtrsim$ 100  & $\sim$ 10$^5$  & $\sim$ 10  & 10$^6$--10$^8$    \\

\hline
\end{tabular}
\end{threeparttable}
\end{table*}

\bsp	
\label{lastpage}
\end{document}